\documentclass[8.5pt,twoside,twocolumn]{article}
\usepackage[super,sort&compress,comma]{natbib} 
\usepackage{mhchem}
\usepackage{times,mathptmx}
\usepackage{sectsty}
\usepackage{balance} 

\usepackage{graphicx} 
\usepackage{lastpage}
\usepackage[format=plain,justification=raggedright,singlelinecheck=false,font=small,labelfont=bf,labelsep=space]{caption} 
\usepackage{fancyhdr}
\begin{document}

\thispagestyle{plain}
\fancypagestyle{plain}{
\fancyhead[L]{\includegraphics[height=8pt]{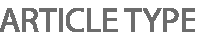}}
\fancyhead[C]{\hspace{-1cm}\includegraphics[height=20pt]{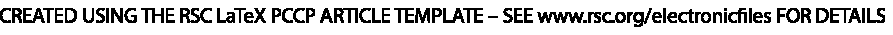}}
\fancyhead[R]{\includegraphics[height=10pt]{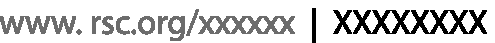}\vspace{-0.2cm}}
\renewcommand{\headrulewidth}{1pt}}
\renewcommand{\thefootnote}{\fnsymbol{footnote}}
\renewcommand\footnoterule{\vspace*{1pt}%
\hrule width 3.4in height 0.4pt \vspace*{5pt}} 
\setcounter{secnumdepth}{5}

\makeatletter 
\def\subsubsection{\@startsection{subsubsection}{3}{10pt}{-1.25ex plus -1ex minus -.1ex}{0ex plus 0ex}{\normalsize\bf}} 
\def\paragraph{\@startsection{paragraph}{4}{10pt}{-1.25ex plus -1ex minus -.1ex}{0ex plus 0ex}{\normalsize\textit}} 
\renewcommand\@biblabel[1]{#1}            
\renewcommand\@makefntext[1]%
{\noindent\makebox[0pt][r]{\@thefnmark\,}#1}
\makeatother 
\renewcommand{\figurename}{\small{Fig.}~}
\sectionfont{\large}
\subsectionfont{\normalsize} 

\fancyfoot{}
\fancyfoot[LO,RE]{\vspace{-7pt}\includegraphics[height=9pt]{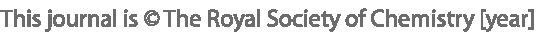}}
\fancyfoot[CO]{\vspace{-7.2pt}\hspace{12.2cm}\includegraphics{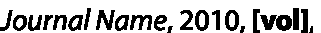}}
\fancyfoot[CE]{\vspace{-7.5pt}\hspace{-13.5cm}\includegraphics{RF}}
\fancyfoot[RO]{\footnotesize{\sffamily{1--\pageref{LastPage} ~\textbar  \hspace{2pt}\thepage}}}
\fancyfoot[LE]{\footnotesize{\sffamily{\thepage~\textbar\hspace{3.45cm} 1--\pageref{LastPage}}}}
\fancyhead{}
\renewcommand{\headrulewidth}{1pt} 
\renewcommand{\footrulewidth}{1pt}
\setlength{\arrayrulewidth}{1pt}
\setlength{\columnsep}{6.5mm}
\setlength\bibsep{1pt}

\twocolumn[
  \begin{@twocolumnfalse}
\noindent\LARGE{\textbf{Shear-Induced Reversibility of 2D Colloidal Suspensions in the Presence of Thermal Noise$^\dag$}}
\vspace{0.6cm}

\noindent\large{\textbf{Somayeh Farhadi $^{\ast}$,
Paulo E. Arratia\textit{$^{a}$}}}\vspace{0.5cm}

\noindent\textit{\small{\textbf{Received Xth XXXXXXXXXX 20XX, Accepted Xth XXXXXXXXX 20XX\newline
First published on the web Xth XXXXXXXXXX 200X}}}

\noindent \textbf{\small{DOI: 10.1039/b000000x}}
\vspace{0.6cm}

\noindent \normalsize{The effects of thermal noise on particle rearrangements in cyclically sheared colloidal suspensions are experimentally investigated, using particle tracking methods. The experimental model system consists of polystyrene particles adsorbed at an oil-water interface, in which the particles exhibit small but non-negligible Brownian motion. We perform experiments on bidisperse (1.0 and 1.2~$\mu$m in diameter) colloidal samples with area fractions $\phi$ of 0.20 and 0.32. We characterize the reversibility of particle rearrangements, and show that unlike dense athermal systems, reversible clusters are not stable; once a particle enters into a reversible trajectory, it has a non-zero probability of becoming irreversible in the following shearing cycle. This probability was previously found to be approximately zero for an analogous athermal system. We demonstrate that the stability of reversibility depends both on packing fraction, $\phi$, and strain amplitude, $\gamma_0$. In addition, similar to previously studied athermal system, we identify hysteresis in the dynamics of rearrangements for reversible particles, which indicates that such reversible rearrangements are dissipative. However, at lower packing fractions, where the dynamics is moved closer to equilibrium by thermal noise, this hysteresis becomes less prominent.  
}
\vspace{0.5cm}
 \end{@twocolumnfalse}
  ]

\section{Introduction}





Particulate systems are ubiquitous in nature and in technology~\cite{larson1999}, and examples include pastes, paints, granular matter, foams, and metallic glasses~\cite{chen2008,yodh2010}. A common feature of these materials is their ability to flow under external load, while maintaining a solid (or jammed) state if unperturbed~\cite{hecke}. Importantly, the macroscopic flow behavior of such particular systems is a strong function of the material microstructure. Understanding the mechanism that governs the dynamics of the material microstructure is crucial in control and process of these materials.  

Colloidal suspensions have been widely used as a model for disordered molecular and atomic systems such as soft and metallic glasses~\cite{berthier2011,royall2015}, where measuring individual particle positions is a challenging task~\cite{sheng2006,chen2010}. In particular, colloidal suspensions can be very useful in the study of microscopic (particle-scale) fluctuations associated with mesoscale rearrangements, which subsequently affect the material properties such as bulk stiffness and plasticity~\cite{royall2015}. Particle fluctuations in colloidal systems are governed by two distinct mechanisms: Brownian (thermal) motion and externally driven deformations (e.g. shear). A number of numerical simulations suggest that an effective temperature exists for dense athermal systems, which sets the energy scale of shear-induced fluctuations~\cite{liu2007,ono2002}. Yet, with some notable exceptions~\cite{chen2010,chikkadi2012}, very few experiments have directly measured thermal fluctuations of colloids in presence of shear. In this study, by shrinking the particle size to $1 \mu m$, where particles exhibit non-negligible Brownian motion, we add a minimal thermal noise to particle dynamics, and study their trajectories under applied cyclic shear.

In a recent study~\cite{keim2013,keim2014,keim2015}, the criteria for reversibility of rearrangements under cyclic shear was comprehensively studied for a dense athermal colloid. In this manuscript, we characterize the effects of thermal noise on the reversibility of particle rearrangements. We probe this effect for two systems with different area fractions ($\phi=0.32$, and $\phi=0.20$), and hence, an order of magnitude difference in their self-diffusivities. This provides us two distinct systems: (i) One system, in which the thermal fluctuation of particles is very small, which provides a high P\'{e}clet regime, where shear is considered as the dominant driving mechanism; and (ii) a system with non-negligible thermal fluctuation of particles, in which both shear and diffusion become important.

Particle rearrangements in colloidal suspensions are traditionally measured either by confocal microscopy or light scattering~\cite{Prasad2007,Dang2016,royall2015,weeks2000,weeks2002_2,weeks2002,berthier2005,petekidis2002} methods. In confocal microscopy, the time resolution of measurements is limited by scanning time, while for scattering the trajectories of individual particles are not available, and instead only the correlated motion of a large group of particles is measured. Here, by using a custom-made interfacial shearing apparatus ~\cite{keim2013,keim2014,keim2015}, we shear and track nearly $4 \times 10^4$ particles adsorbed at an oil-water interface, and characterize reversibility, as well as the onset of rearrangement, for each particle. On the other hand, stroboscopic reversibility gives us useful information on the fabric of configurational energy landscape. In particular, it manifests the existence of energy metabasins which restrict the dynamics of the system. Our data shows that the dynamics within these metabasins is strongly affected in the presence of small thermal noise.

\section{Experimental Methods}
\begin{figure}
\centering
{\includegraphics[scale=0.13]{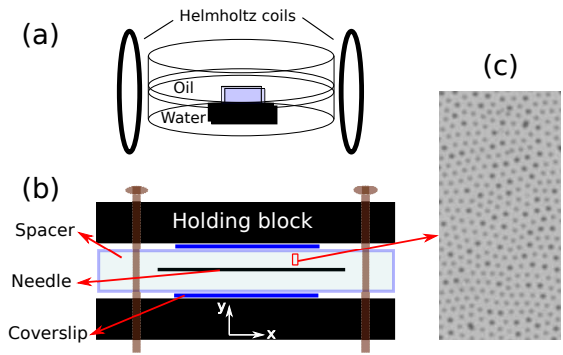}}
\caption{a) Schematics of the shearing cell. b) Schematics of the spacer apparatus (top view). c) Mixture of $1.0$ and $1.2 \mu$m particles visualized under microscope. The area fraction is $\phi=0.20$.
}
\label{fig:schematics-rheology}
\end{figure}

The effects of thermal noise on particle rearrangements in colloidal suspensions are investigated using a custom-made interfacial shearing cell~\cite{Reynaert2007,brooks1999}. In this apparatus, the particle suspension is confined to a monolayer, which allows for the visualization of the evolution of the fluid microstructure by tracking individual particles. As shown in Fig.~\ref{fig:schematics-rheology}(a,b), a thin magnetized needle is embedded at the oil-water interface to be studied, inside an open channel formed by 2 walls. An electromagnet forces the needle, creating a uniform shear stress on the material between the needle and the walls. The region from which the data is acquired is near the center of the channel and sufficiently far from both tips of the needle, which avoids the non-uniform flow around the needle tips (more details below and elsewhere~\cite{keim2013,keim2014,keim2015}).

In order to create a well-controlled shearing flow, a small device is built to hold and control the distance between two microscope cover-slips that serve as parallel walls (Fig.~\ref{fig:schematics-rheology}b). The device is placed inside a $10$~cm diameter Petri dish and then partially filled with DI water. A small amount of oil, decane 99+\% from ACROS, is then carefully poured on top of the water, creating a thin oil layer and an (oil-water) interface between the two coverslips. The magnetized needle is made from phosphate-coated carbon steel wire (from Mcmaster-Carr) and is $4$~cm long, 0.15~mm in diameter, and 4 mg in mass. Prior to preparation, all of the parts are sonicated and rinsed with DI water, followed by a rinse with ethanol to avoid any aggregation inducing contaminations. 

Next, we inject particles to the oil-water interface. Due to the high surface tension, particles are adsorbed and trapped at the interface, forming a stable particle monolayer \cite{keim2013,keim2014,keim2015}. This monolayer is disordered because particle size distribution is bidisperse, and also the charge distribution around each particle is non-uniform that results in asymmetrical particle-particle interaction.
The magnetic needle is then carefully placed at the monolayer, and the needle's weight is supported by capillary forces. 
A static Helmholtz field keeps the needle centered in the channel, while additional electromagnets move the needle back and forth, uniformly shearing the interface in the channel. The schematics of our custom made setup is shown in Fig.~\ref{fig:schematics-rheology}. 


The colloidal particles used in this experiment are sulfate latex beads (8\% w/v, Invitrogen). These microspheres are charge-stabilized in DI water due to their surface sulphate treatment. Once placed at the interface, they form dipoles with a long-range and repulsive interaction force. We use equal volumes of 1~$\mu$m and 1.2~$\mu$m particles to form bidisperse colloidal mixtures. Here, bidispersity is used to model a disordered system, by preventing crystallization, and consequently, long-range ordering effects. We dilute $0.1$mL of the mixture to $1$mL with DI water, and add $0.5$mL isopropanol (Fisher Scientific) for easy dispersion at the interface. The particle mixtures are then slowly injected to the interface using a micro-pipet. More details on the preparation protocol is provided in references \cite{keim2013,keim2014,keim2015}. By injecting different volumes of particle suspension, we are able to control the area fraction at the interface. In this study, we provide data for two area fractions: $\phi=0.20$, and $\phi=0.32$. The area fractions are calculated as $\phi=\frac{A_p}{A}$, where $A$ is a sampling region, and $A_p$ is the area spanning by particles in that region. The particle monolayer is imaged with a long-distance inverted microscope (K2/SC Infinity Photo-Optical) and high-speed camera (Flare 4M180, IOIndustries); data is taken at $40$, $60$, or $80$ fps. Within the recorded images (Fig.~\ref{fig:schematics-rheology}c), $1\mu$m particles approximately span 6 pixels. We keep the temperature constant ($23\,^{\circ}\mathrm{C}$) for all of the presented data, by confining the sample Petri dish in a glycerol-filled bath.

For all experiments, we keep the needle frequency constant at  $f=$0.1Hz. We have chosen 0.1Hz in our experiments because we want to maintain a linear response from the clean interface (before adding particles). In particular, we measure $G^*=|G'+iG''|$, where $G'$ and $G''$ are respectively the elastic and loss moduli of the interface, and identify a region where $G^*$ is independent of frequency \cite{keim2014}. For our system, we find that $G^*$ begins to rapidly grow at approximately 0.2Hz, which sets our upper limit for frequency. Also, as we move to lower frequencies below 0.05Hz, we begin to experience ambient vibrational noise. Due to these two issues, we chose 0.1Hz as our fixed experimental frequency. To consistently prepare the material for each experiment and to avoid memory effects, oscillatory forcing at large strain amplitude ($\gamma_0=1.0$) is initially performed for 3 cycles and then stopped. In the experiments presented here, the values of the strain amplitude ($\gamma_0$) range from 0.01 up to 0.2. A sample video of the experiment is provided in the Supplementary Materials. The images are then processed using {\it trackpy}, a python based particle tracking package\cite{trackpy}. For each data set, nearly $4\times10^4$ particles are identified and tracked. Particle trajectories are recorded up to 60 cycles.

\section{Analysis}
\begin{figure}
\centering
{\includegraphics[scale=0.54]{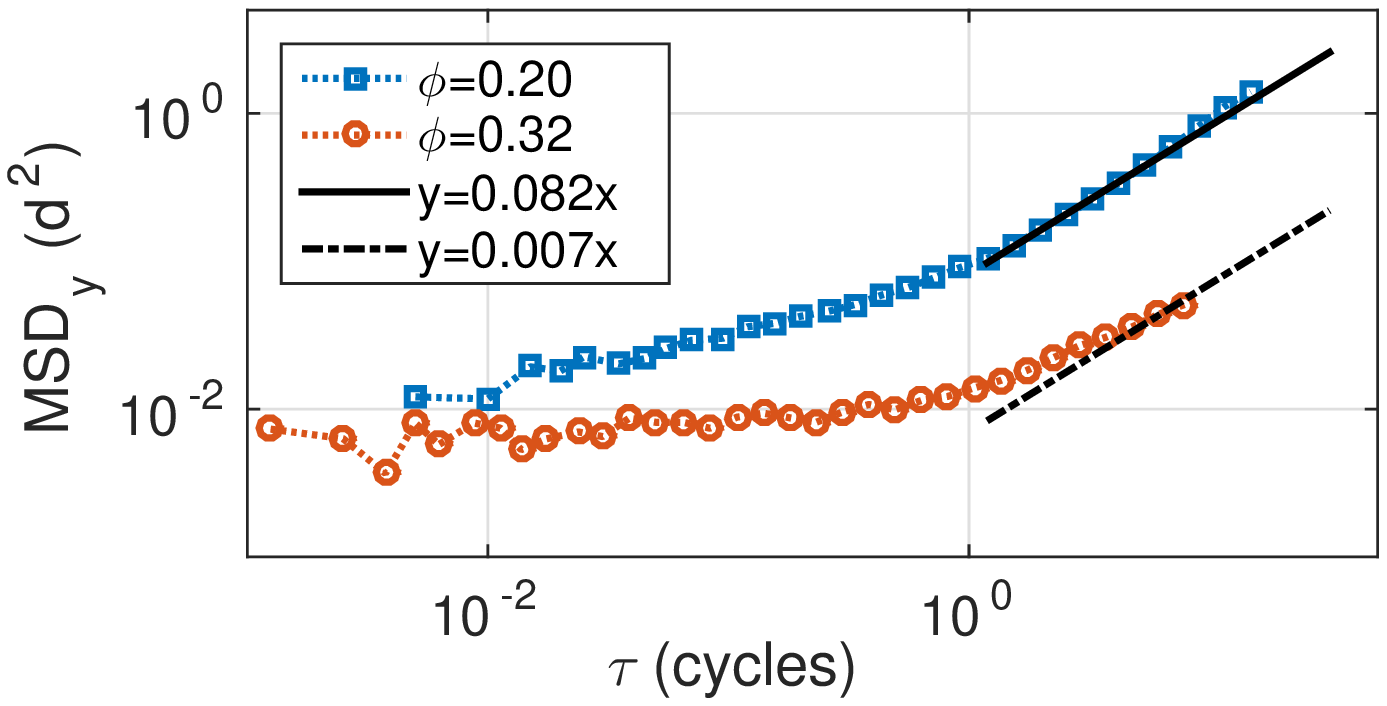}}
{\includegraphics[scale=0.54]{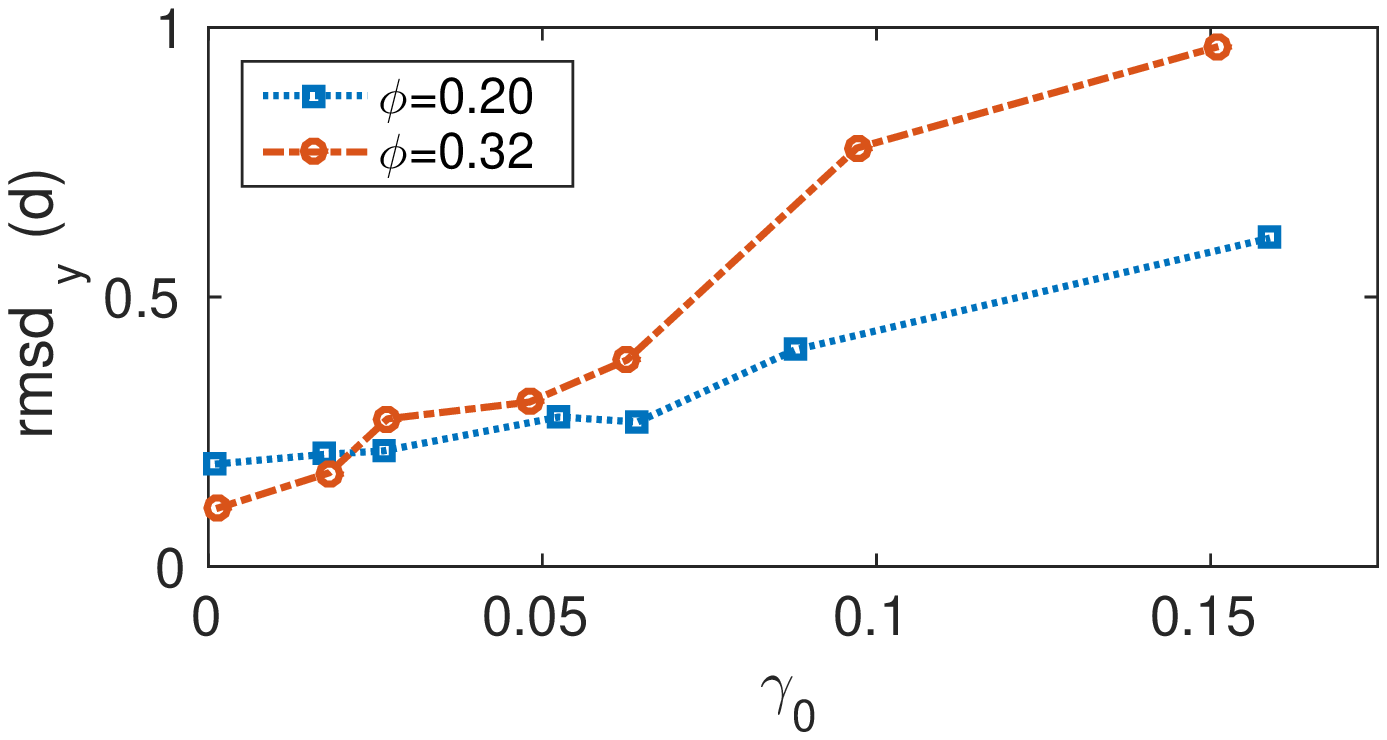}}
\caption{a) Mean squared displacement of particles in quiescent state, measured for transverse to shear direction (from the y component of displacements). b) Root mean squared displacement in y direction, transverse to shear. Here, $rmsd_y$s are calculated for a time equivalent of one shearing cycle. The length unit, d, is 1$\mu m$, or the diameter of small particles, and the time unit is one shearing cycle ($10 s$). }
\label{fig:alpha}
\end{figure}

We begin our analysis of thermal effects on particle rearrangements by characterizing particle diffusivity in the sample. We use particle mean squared displacement, $MSD_y(\tau)=<(y(t+\tau)-y(t))^2>$, to characterize the particle diffusion in the quiescent states (i.e., in the absence of shear). Here, $y$ is the component of particle position perpendicular to the wall, and $\tau$ is the time-lapse. Fig.~\ref{fig:alpha}a shows the $MSD_y$ as a function of shear cycle $\tau$ for colloidal suspensions of volume fraction 0.20 and 0.32; as expected, particles in the dilute suspension ($\phi=0.20$) tend to have larger mean displacements compared to the dense suspension ($\phi=0.32$). We use the long-time data where a linear regime is observed to extract the sample diffusivities (see linear fits in Fig.~\ref{fig:alpha}a) using the relationship $MSD_y=2 D_{\infty} \tau$. We find that $D_{\infty}=0.0035 d^2/c= (0.00035 \mu m^2/s)$ for $\phi=0.32$, and $D_{\infty}=0.041 d^2/c= (0.0041 \mu m^2/s)$ for $\phi=0.20$, where $d$ is the particle diameter and $c$ is the time scale of one shearing cycle.

The relative importance of convection (or flow) to diffusion in a system is usually characterized by the P\'{e}clet number. We define a modified P\'{e}clet number for sheared states as $Pe^{\star}=\overline{|\dot{\gamma}|}d^2/2 D_{\infty}$, which is the ratio of the flow time scale, $\overline{|\dot{\gamma}|}$, and diffusion time scale, $\frac{2 D_{\infty}}{d^2}$. Here, $d$ is the particle diameter, and $\overline{|\dot{\gamma}|}$ is the strain rate magnitude averaged over a complete shear cycle. For a sinusoidal strain imposed by the needle, $\gamma=\gamma_0 sin(\frac{2\pi}{T} t)$, the average shear rate of the needle over one cycle is $\overline{|\dot{\gamma}|}=\frac{4\gamma_0}{T}$. This gives the range of modified P\'{e}clet numbers for the probed strain amplitudes presented in this manuscript as: $0.8<Pe^{\star}<7.8$ for $\phi=0.20$, and $10.3<Pe^{\star}<86.3$ for $\phi=0.32$.
While $\phi=0.32$ could be considered in an athermal regime, for $\phi=20$, the thermal effects are non-negligible.


Also note that $MSD/-y$ for $\phi=0.20$ monotonically increases vs. time-lapse, which signifies a weak caging effect ($\alpha$-relaxation). For the high packing fraction ($\phi=0.32$), $MSD_y$ starts to plateau before reaching its diffusive regime. This plateau indicates that caging starts to dominate the dynamics, and at $\phi=0.32$, the system is close to its glass transition point~\cite{weeks2002_2,weeks2000}.

Next, we quantify particle diffusivity in the lateral direction in the presence of shear by computing the root-mean-square displacement $rmsd_y$ for a time scale equivalent to one shearing cycle (i.e. $\frac{1}{f}=10s$). Fig.~\ref{fig:alpha}b shows the quantity $rmsd_y$ as a function of strain amplitude $\gamma_0$ for the $\phi=0.20$ and $\phi=0.32$ samples. We find that the values of $rmsd_y$ increases monotonically with strain amplitude for both samples. However, the increase is much weaker for the system with lower packing fraction ($\phi=0.20$). That is, particles in the $\phi=0.32$ sample show larger displacements under shear than a more dilute sample; in the limit of very low shear (or strain amplitude), the opposite is observed.

\begin{figure}
\centering
{\includegraphics[scale=0.5]{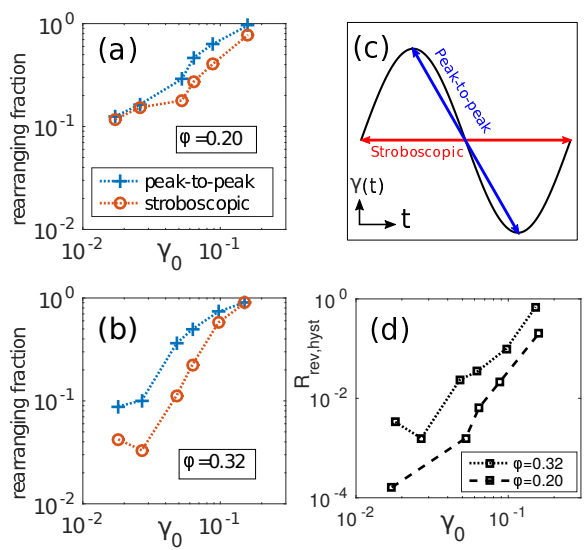}}
\caption{Fraction of stroboscopic and peak-to-peak rearranging particles for (a) $\phi=0.20$, and b) $\phi=0.32$. c) Fraction of hysteretic reversible particles ($R_{rev,hyst}$).d) schematics of peak-to-peak and stroboscopic sampling.}
\label{fig:rearranging_fraction}
\end{figure}

A key advantage of this interfacial shear cell is that one can obtain detailed imaging of colloidal suspension microstructure while undergoing cyclic shear. We will now focus on stroboscopic rearrangements, which we define as particles that change neighbors after completing a shearing cycle. The change in neighbors is measured stroboscopically by sampling at times $t_n= t_{0}+ n 2\pi \omega^{-1}$, $n=\{0,1,2,...\}$, so that we compare the beginning and end of a full period of driving that straddles $t_{n}$ and $t_{n+1}$ (Fig.~\ref{fig:rearranging_fraction}c). Particle rearrangements are characterized by $D^2_{min}=\Sigma(r'_i-\textbf{E}r_i)^2$, which quantifies the mean-squared deviation of particle displacements from the best-fit affine deformation of particles during a time interval $\Delta t$~\cite{falk1998,liu2014}. Sufficiently high values of $D^2_{min}$ denote rearrangements~\cite{cande2010}. The association of high values of $D^2_{min}$ with rearrangements enables us to identify the events in which particles change neighbors, and in other words, break their cages. For a given particle $p$, $r'_i$ and $r_i$ are the positions of its neighbouring particles before and after applying a deformation. Usually, the neighbouring group of particles is chosen as all of the particles centered within 2.5 $\sigma_{aa}$  relative to the center of particle $p$, where $\sigma_{aa}$ is the distance corresponding to the first peak of pair correlation function, $g(r)$. Note that $\sigma_{aa}$, which represents the average interparticle spacing, becomes larger as the packing fraction, $\phi$, is reduced. $\textbf{E}$ is the least squares fit, which transforms $r_i$ to $r'_i$ affinely. With this definition, $D^2_{min}$ is a measure of nonaffinity of particle displacements, and its value quantifies the extent of rearrangement for a given particle.

\begin{figure}
\centering
{\includegraphics[scale=0.35]{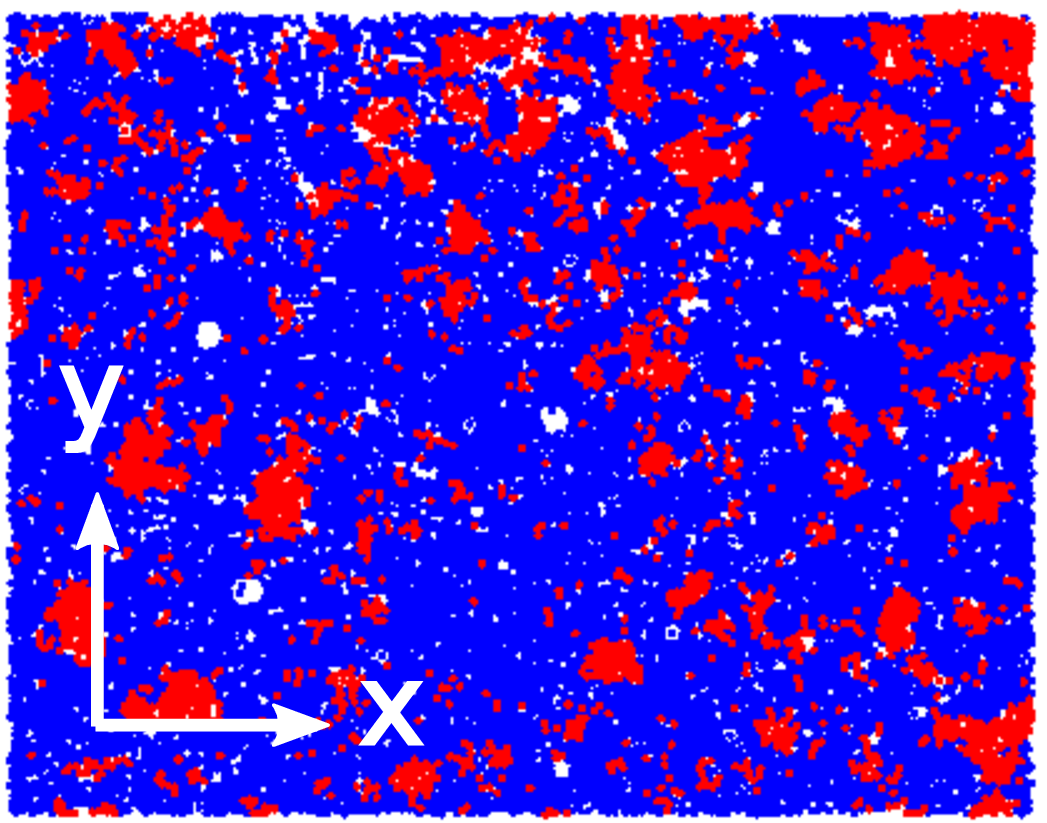}}
\hspace{3mm}
{\includegraphics[scale=0.35]{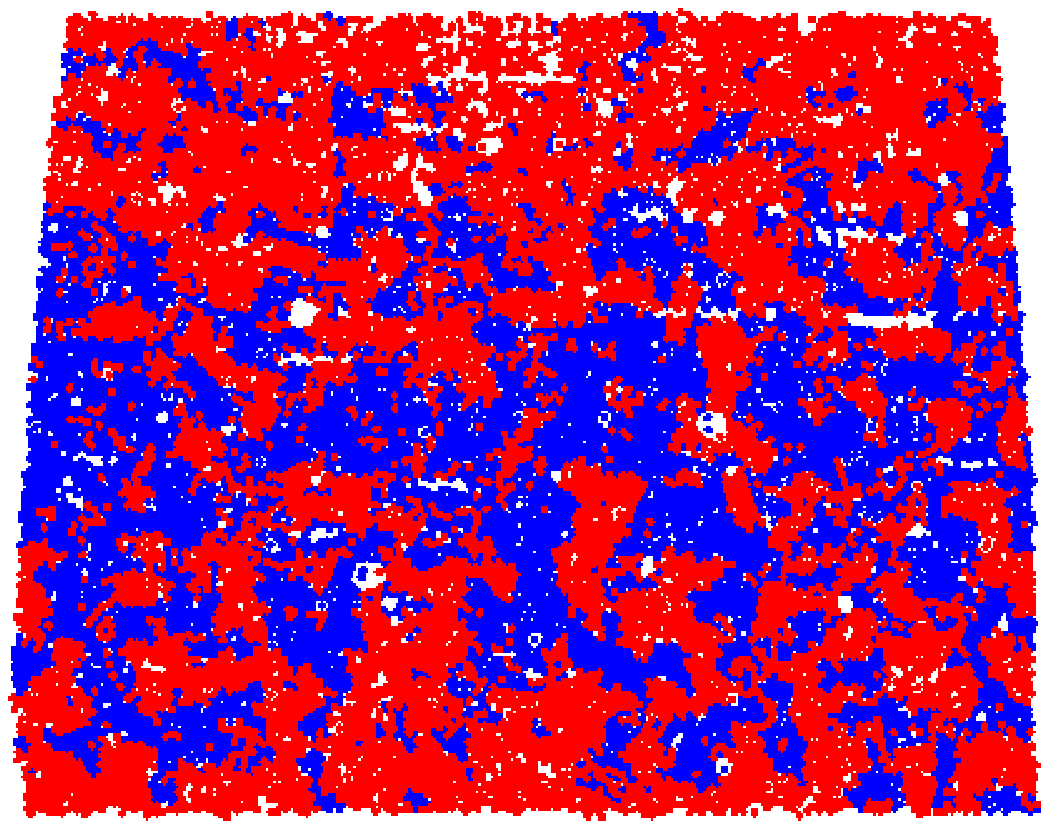}}
\caption{Rearranging (Red), and nonrearranging (blue) particles for $\gamma_0=0.017$ (left), and $\gamma_0=0.158$ (right). The packing fraction is $\phi=0.20$, and both distributions are plotted for cycle 15. A fraction of particles are still reversible, even for a relatively high strain value of $\gamma_0=0.158$. The rectangular region demonstrates a 0.5$\times$0.3mm sample. The flow is in x direction.}
\label{fig:rearranging_area}
\end{figure}

Fig.~\ref{fig:rearranging_area} shows snapshots of the spatial distribution of stroboscopically rearranging particles for a colloidal suspension of volume fraction $\phi=0.20$ for $\gamma_0=0.017$ (Fig.~\ref{fig:rearranging_area}a) and $\gamma_0=0.158$ (Fig.~\ref{fig:rearranging_area}b). Particles colored in red undergo irreversible rearrangements while particle colored in blue undergo reversible motion (i.e., no stroboscopic rearrangements).  Here, we take $D^2_{min}>0.015$ as the threshold for rearrangement (this threshold is set by the noise level in our $D^2_{min}$ calculation). For low strain amplitude ($\gamma_0=0.017$, Fig.~\ref{fig:rearranging_area}a), we find a relatively small number of rearranging particles (red regions) that form regions of scattered clusters. As the strain amplitude is increased ($\gamma_0=0.158$, Fig.~\ref{fig:rearranging_area}b), we find many more rearranging particles. However, we still see a considerable fraction of particles that are not stroboscopically rearranging and are reversible (blue regions).

\begin{figure}
\centering
{\includegraphics[scale=0.57]{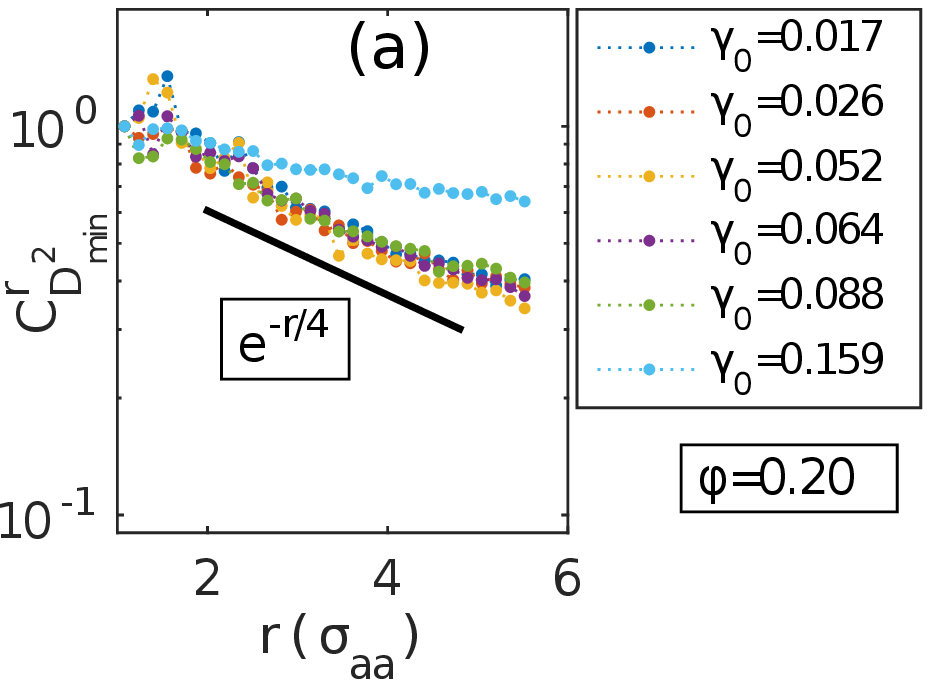}}
{\includegraphics[scale=0.57]{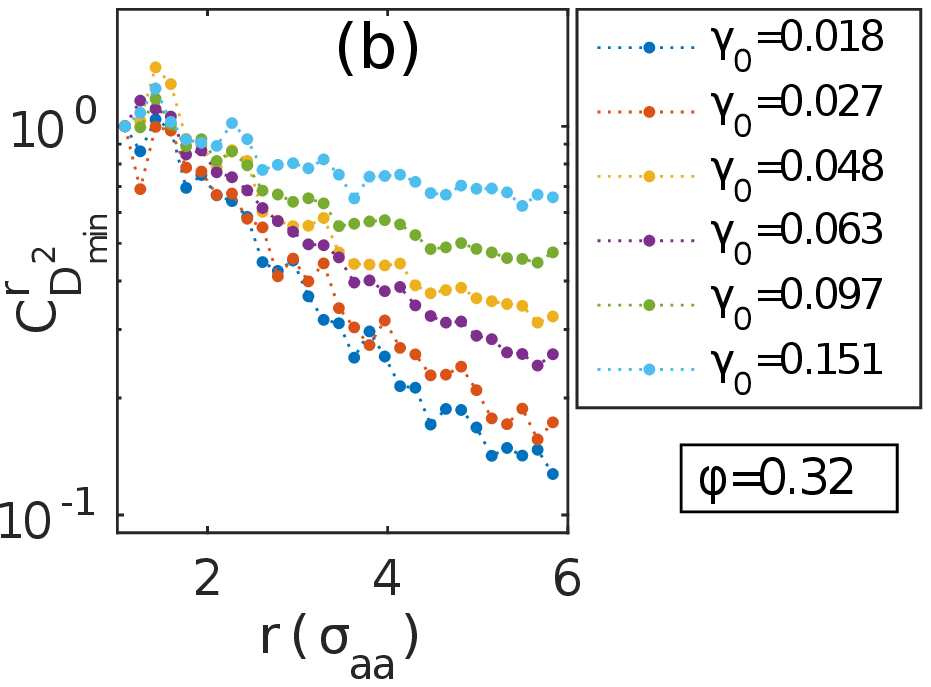}}
\caption{Spatial correlation of $D^2_{min}$ ($C^r_{D^2_{min}}$)for $\phi=0.20$ and $\phi=0.32$. For the dilute system, the correlation length remains almost constant as $\gamma_0$ is increased. $\sigma_{aa}$ is the length scale associated with the first peak of $g(r)$, which quantifies the  average interparticle spacing for each system.}
\label{fig:rearranging_corr}
\end{figure}

In order to gain insight into the relationship between particle rearrangements and the imposed shear deformation, we compute the spatial correlation of $D^2_{min}$ ($C^r_{D^2_{min}}$) for particles undergoing irreversible rearrangements (colored red). The quantity $C^r_{D^2_{min}}$ for different values of strain amplitude $\gamma_0$, for $\phi=0.20$ and $\phi=0.32$, are shown in Fig.~\ref{fig:rearranging_corr}a and Fig.~\ref{fig:rearranging_corr}b, respectively. For $\phi=0.20$, the correlation length remains nearly identical as $\gamma_0$ is increased, except for the largest amplitude case. However, for $\phi=0.32$, the correlation length is highly dependent on the strain amplitude. This data shows that as we move to a denser state, stroboscopic rearrangements are mostly governed by external driving (i.e. shear), and are less affected by thermal noise. Similar behavior was previously observed in simulations as the system was moved from shear dominated to thermal dominated regime~\cite{Chikkadi2012}.

We have shown that even at large strain amplitudes, there are still quite a few stroboscopically reversible particles or reversible rearrangements (Fig.~\ref{fig:rearranging_area}b). Do these particles undergo rearrangements within a given cycle? Here, we investigate the existence and statistics of such rearrangements by measuring the quantity $D^2_{min}$ for rearrangements between the two peaks of the sinusoidal strain cycle, or the ``peak-to-peak $D^2_{min}$''. For this, we study particle positions at times $t_n= t_{0}+ \frac{n}{2}\pi \omega^{-1}$, $n=\{1,3,5,...\}$ (Fig.~\ref{fig:rearranging_fraction}c). We find that (for all experiments presented here) a fraction of particles which appear stroboscopically reversible, do rearrange ($D^2_{min}>0.015$) when viewed from peak-to-peak. Fig.~\ref{fig:rearranging_fraction}(a,b) show the fraction of rearrangement type, stroboscopic versus peak-to-peak, for $\phi=0.20$ and $\phi=0.32$ as a function of strain amplitude. We find that for most cases, the fraction of particles undergoing peak-to-peak rearrangement is larger than stroboscopic rearrangement; that is, a significant number of particle rearrangements are reversible. As the strain amplitude is increased, the fraction of particles undergoing both types of rearrangements also increase. 
Note that the fractions of reversible particles (fraction of stroboscopic rearrangements  subtracted from fraction of peak-to-peak rearrangements) are relatively smaller for $\phi=0.20$ compare to $\phi=0.32$ (Fig.~\ref{fig:rearranging_fraction}a,b). Also, the peak-to-peak rearrangements behave very similarly for both packing fractions, while the stroboscopic rearrangements are highly dependent on the packing fraction, particularly for the low strain amplitudes. This suggests that, while cyclic driving at small strain amplitudes self-organizes the system to reversible states, thermal noise counter-affects the organization, and makes the dynamics irreversible.  

Previous studies on a jammed athermal system, identified non-rearranging (reversible) particles which were dissipating energy, and were referred to as `plastic reversible'~\cite{keim2013,keim2014,keim2015}. A plastic reversible particle undergoes a rearrangement which completely reverses within a strain cycle, yet the rearrangement path has hysteresis. We now investigate whether the reversible rearrangements found here are also plastic. 
In order to identify hysteresis, we define the strain at which a given particle starts to rearrange as $\gamma^{on}$, and the strain at which the particle stops rearranging as $\gamma^{off}$.
Consider a strain cycle $\gamma(t)$ (Fig.~\ref{fig:rearranging_fraction}c). We then take the $\gamma(t=0)$ as the reference frame, and calculate $D^2_{min}$ values for all of the frames in the cycle with respect to this reference frame. Using the threshold $D^2_{min}=0.015$ as the threshold for rearrangement, we obtain $\gamma^{on}>\gamma(0)$ as the last strain where $D^2_{min}<0.015$, and $\gamma^{off}$ as the first strain where $\gamma^{off}>\gamma^{on}$, and $D^2_{min}<0.015$. We require $\gamma^{on}$ and $\gamma^{off}$ to be in the first and second halves of the $\gamma$ function. Note that for a hysteretic particle, $\gamma^{on}-\gamma^{off}\neq0$.

Fig.~\ref{fig:rearranging_fraction}d shows the fraction of reversible particles which exhibit hysteresis in their rearrangement path as a functions of strain amplitude. We quantify hysteresis in the sample using the quantity $R_{rev,hyst}=\frac{N_{rev,hyst}}{N_{rev}}$, where, $N_{rev}$ is the number of all reversible particles and $N_{rev,hyst}$  is the number of all reversible particles which exhibit hysteresis in their rearrangement path. We have excluded the first 3 transient cycles from the statistics. The data clearly indicate the existence of plastic reversible paths even in the presence of thermal noise. However, the fraction of plastic reversible particles is considerably smaller for $\phi=0.20$ compared to $\phi=0.32$. And, although irreversible particles are more frequent for $\phi=0.20$, the hysteresis of rearrangements for its reversible particles is significantly lower than for $\phi=0.32$. A possible explanation of this behavior is that, since the thermal fluctuations are stronger at lower packing fractions, the dynamics is closer to equilibrium, and hence, non-dissipative.  Similar to a glass forming liquid, as the packing fraction is lowered below glass point, the cages are broken, and the particles escape the arrested dynamics and diffuse. This diffusive process allows the system to explore a larger region in configurational energy landscape, and consequently, moves the system closer to equilibrium. By increasing the packing fraction, thermal equilibration is hindered and the system regains out-of-equilibrium and athermal properties.

For dense athermal systems, once a particle enters into a reversible cycle it remains reversible in the upcoming cycles\cite{keim2014}. In contrast, thermal noise allows the particles to escape reversible cycles with a certain probability (see Supplementary Materials). Fig.~\ref{fig:rearranging_prob}a shows the transition probabilities (TP) between the reversible/irreversible states for $\phi=0.20$, and $\phi=0.32$ as a function of strain amplitude. The TPs are measured by counting the number of particles transitioning from a particular state at cycle $t$ to another state at cycle $t+1$, and averaged over all cycles. We exclude 3 initial cycles to avoid transient behavior (TPs do not change significantly over non-transient cycles). The notation is as follows: $p[I^{t+1}|R^t]$ is the transition probability (TP) of a particle undergoing reversible rearrangement at time $t$ to switch to an irreversible rearrangement at time $t+1$; similarly, $p[R^{t+1}|I^t]$ is the transition probability (TP) of a particle undergoing irreversible rearrangement at time $t$ to switch to an reversible rearrangement at time $t+1$. Results show that, for both volume fractions, the probability of a particle displaying irreversible rearrangement increases as strain amplitude is increased. The stability of reversible states can be quantified by $S=p[R^{t+1}|I^t]-p[I^{t+1}|R^t]$ (Fig.~\ref{fig:rearranging_prob}b). 
We find that reversible cycles at lower values of $\gamma_0$ and irreversible cycles at higher $\gamma_0$s are more stable. However, we also observe that at lower packing fraction, where thermal noise is significant, the stabilities of both reversible and irreversible particles (respectively, at lower and higher $\gamma_0$) are hindered in comparison with the denser system. This is consistent with our earlier observation (Fig.~\ref{fig:rearranging_corr})
that the dynamics of reversibility is less affected by external driving as we increase the thermal noise in the system.

\begin{figure}
\centering
{\includegraphics[scale=0.2]{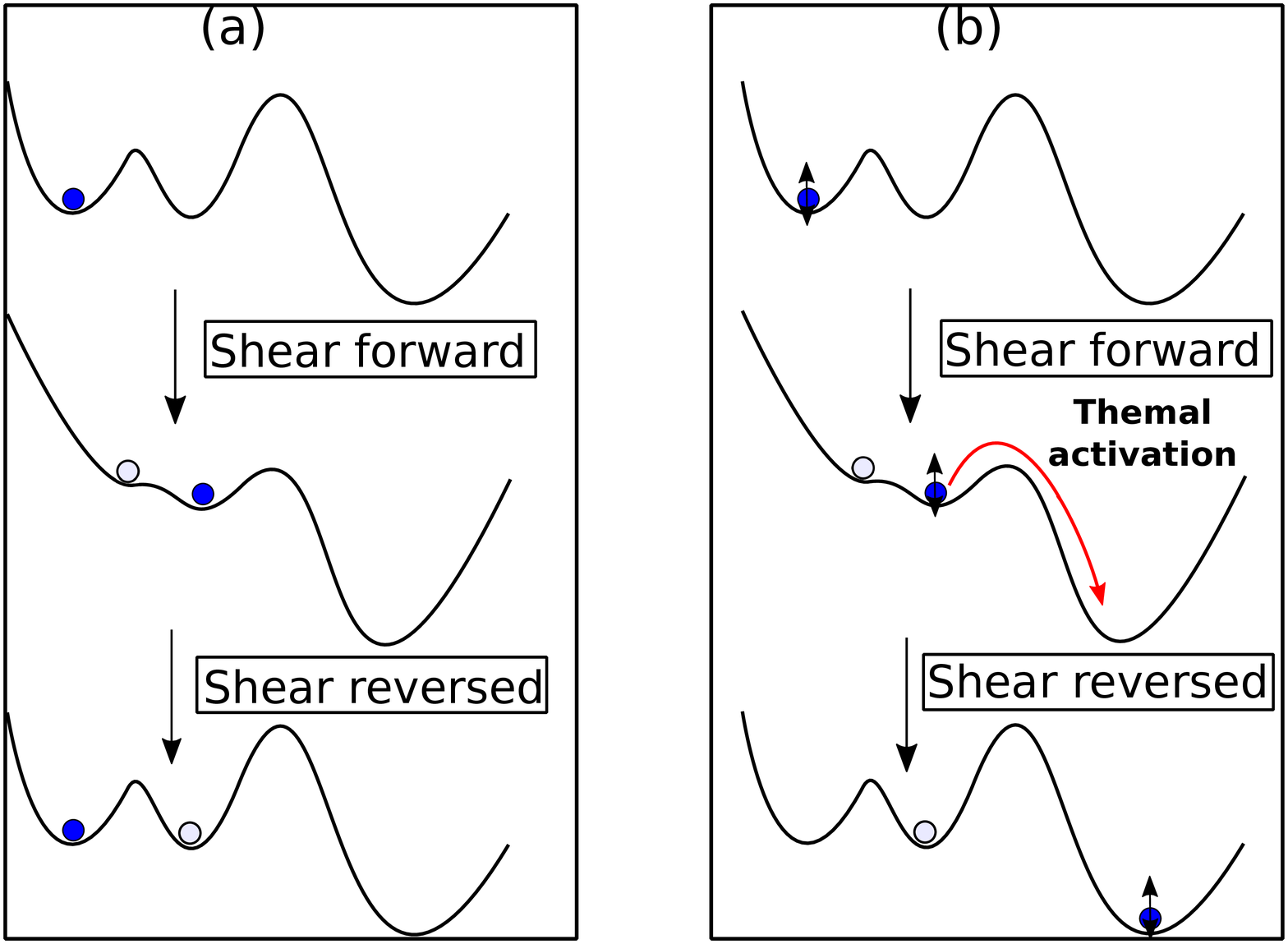}}
\caption{ Schematics of the state dynamics as a result of shear induced activation for a) athermal, and b) thermal systems. }
\label{fig:EL}
\end{figure}

One can understand the reversible-irreversible TPs within configurational energy landscape framework. Initially, a local energy minimum is occupied (Fig.~\ref{fig:EL}a). Application of shear distorts and flattens the local minimum such that the system jumps to a nearby minimum state \cite{lacks2001,Regev2015}. For athermal particles, by reversing shear, the energy landscape resumes its original shape, and the system jumps back to the initial local minimum (reversible regime). This implies the existence of a metabasin in the energy landscape which confines the dynamics to a small region in configurational space\cite{ediger2012,Anderson2002,Regev2015,Heussinger2014}. By increasing the shear amplitude, the distortion of the energy landscape is large enough, such that the system escapes from the metabasin and the trajectories become irreversible. If, in addition to shear, the particles are also thermally activated, the system has a finite probability to escape the metabasin (Fig.~\ref{fig:EL}b). As the thermal energy increases, the probability of escaping the metabasin, and hence irreversibility, also increases.

\begin{figure}
\centering
{\includegraphics[scale=0.4]{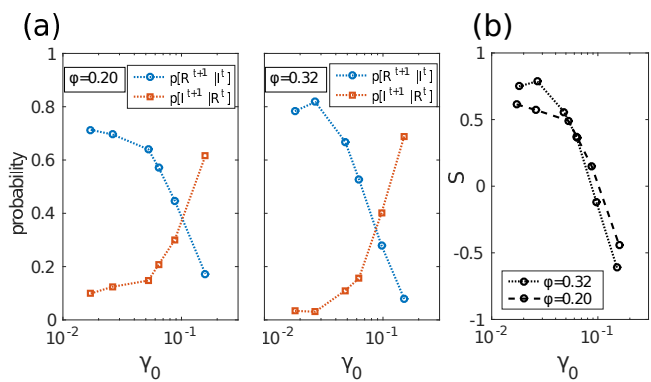}}
\caption{ a) Transition probabilities between reversibility states as a function of strain amplitude, $\gamma_0$. Note that the probability of irreversible rearrangement increases as $\gamma_0$ is increased. b) $S_{r}=p[R^{t+1}|I^t]-p[I^{t+1}|R^t]$ quantifies the stability of reversible states.}
\label{fig:rearranging_prob}
\end{figure}

\section{Conclusions}
In this manuscript, we presented an experimental investigation on particle rearrangements of a 2D colloidal system under cyclic shear in the presence of small but finite thermal noise.  To our knowledge, this is the first experimental study that directly measures thermal fluctuations under shear for a large number of particles ($4\times 10^4$) and for relatively long times. Experiments are performed using a custom-made interfacial shearing cell, and colloidal samples with area fractions $\phi=0.20$ and $\phi=0.32$. The thermal noise in the colloidal suspensions was quantified by the diffusion constant, $D_{\infty}$, extracted from the long time mean squared displacements (Fig.~\ref{fig:alpha}). 

The relative importance of thermal motion to applied flow in our system is quantified using the modified P\'{e}clet number, which ranges from 0.8 to approximately 80 and straddles the thermal and athermal regimes. We find that minimal levels of thermal noise, which is usually negligible in steady shear, has non-negligible effects on the reversibility of particle rearrangements under cyclic driving even for relatively large $Pe^{\star} (\sim 10$). Recently, an investigation by Khair~\cite{khair2016} showed that the instant vanishing of flow at turning points, which moves the system towards low P\'{e}clet regime, could affect the rheology of particle suspensions in strongly nonlinear flows. This vanishing P\'{e} at turning points (i.e. oscillation peaks), and hence the dominance of diffusion over shear, could also be a possible explanation for the dynamics we observe, specifically the fraction and the instability of reversible rearrangements, compared to an athermal system. We also show that even though the thermal noise is slightly lower in the quiescent state of the denser system ($\phi=0.32$), as the shear amplitude (and consequently the average shear rate) is increased, the shear induced fluctuations grow faster compared to the more dilute system ($\phi=0.20$).


Particle rearrangements are characterized using the quantity $D^2_{min}$, which quantifies the mean-squared deviation of particle position at time $t$ from the best-fit affine deformation of its neighborhood during a time interval $\Delta t$~\cite{falk1998,liu2014}. Rearrangements are characterized by having high $D^2_{min}$ values. We find that the correlation length of rearrangements remains almost constant for the low packing fraction case $\phi=0.20$, while for $\phi=0.32$ the correlation length is strongly dependent on strain amplitude. This indicates that, for higher packing fractions, thermal effects are hindered and shear is the dominant driving mechanism for rearrangements. Further analysis of particle rearrangements showed that, similar to dense athermal systems, a fraction of particles undergo plastic reversible cycles, which decrease in number as the packing fraction decreases. That is, plastic reversibility is diminished as we move the system (slightly) towards thermal equilibration, here by lowering $\phi$. We also observed that, for thermally activated systems, unlike jammed athermal systems, the particles escape from reversible states with a certain probability which depends both on packing fraction and strain amplitude. We find that for sufficiently small strain amplitudes, the reversible cycles are relatively more stable for higher packing fractions. This suggests that as the liquid is moved close to its glass transition point, the combination of thermal energy and shear is not sufficient to overcome the energy barrier and rearrange the particles towards equilibrium.

We would like to thank N. Keim, A. Yodh, and A. Liu for helpful discussions and suggestions, as well as Madhura Gurjar for the experimental setup design. This work was partially supported by NSF-Penn-MRSEC-DMR-1120901. Acknowledgment is made to the Donors of the American Chemical Society Petroleum Research Fund for partial support of this research.


\begin{thebibliography}{9}




\bibitem{larson1999} R. G. Larson, The rheology and structure of complex fluids. Oxford University Press: New York, 1999. 

\bibitem{yodh2010} D.T. Chen, Q. Wen, P.A. Janmey, J.C. Crocker, A.G. Yodh,  Annu. Rev. Condens. Matter Phys. {\bf 1}, 301-322 (2010).

\bibitem{chen2008} M. Chen,  Annu. Rev. Mater. Res. {\bf 38}, 445-469 (2008).

\bibitem{hecke} M. Van Hecke, J. Phys. Condens. Matter {\bf 560}, 1-75 (2015).

\bibitem{royall2015} C. Royall, S. Williams, Phys. Rep. {\bf 560}, 1-75 (2015).

\bibitem{berthier2011} L. Berthier, G. Biroli, Rev. Mod. Phys. {\bf 83}, 587 (2011).

\bibitem{sheng2006}H.W. Sheng, W.K. Luo, F.M. Alamgir, J.M. Bai, E. Ma, Nature {\bf 439}, 419-425 (2006).

\bibitem{chen2010} D. Chen, D. Semwogerere, J. Sato, V. Breedveld, E.R. Weeks, Phys. Rev. E {\bf 81}, 011403 (2010).

\bibitem{liu2007} T.K. Haxton, A.J. Liu, Phys. Rev. Lett. {\bf 99}, 195701 (2007).


\bibitem{ono2002} I.K. Ono, C.S. O’Hern, D.J. Durian, S.A. Langer, A.J. Liu, S.R. Nagel, Phys. Rev. Lett. {\bf 89}, 095703 (2002).


\bibitem{chikkadi2012} V. Chikkadi, P. Schall, Phys. Rev. E {\bf 85}, 031402 (2012).

\bibitem{Leutheusser1984} E. Leutheusser, Phys. Rev. A {\bf 29}, 2765 (1984).

\bibitem{Berthier2002}L. Berthier, J.L. Barrat, Phy. Rev. Lett. {\bf 89}, 095702 (2002).

\bibitem{Yamamoto1998} R. Yamamoto, A. Onuki, Phys. Rev. E {\bf 58}, 3515 (1998).

\bibitem{keim2013} N.C. Keim and P.E. Arratia, Soft Matter {\bf 9}, 6222 (2013).

\bibitem{keim2014} N.C. Keim, P.E. Arratia, Phys. Rev. Lett. {\bf 112}, 028302 (2014).

\bibitem{keim2015} N.C. Keim and P.E. Arratia, Soft Matter {\bf 11}, 1539 (2015).

\bibitem{weeks2002} E.R. Weeks, D.A. Weitz, Phys. Rev. Lett. {\bf 89}, 095704 (2002).

\bibitem{Dang2016} M.T. Dang, D. Denisov, B. Struth, A. Zaccone, P. Schall, Eur. Phys. J. E {\bf 39}, 44 (2016).

\bibitem{berthier2005} L. Berthier, G. Biroli, J.-P. Bouchaud, L. Cipelletti, D. El Masri, D. L'H\^{o}te, F. Ladieu, M. Pierno, Science {\bf 310}, 1797 (2005).

\bibitem{weeks2002_2} E. Weeks, D. Weitz, Chem. Phys. {\bf 284}, 361 (2002).

\bibitem{Prasad2007} V. Prasad, D. Semwogerere, E.R. Weeks, J. Phys. Condens. Matter {\bf 19}, 113102 (2007).

\bibitem{weeks2000} E.R. Weeks, J.C. Crocker, AC. Levitt, A. Schofield, D.A. Weitz, Science {\bf 287}, 627 (2000).

\bibitem{petekidis2002} G. Petekidis, A. Moussaïd,
P. N. Pusey, Phys. Rev. E {\bf 66}, 051402 (2002).

\bibitem{trackpy}trackpy: http://dx.doi.org/10.5281/zenodo.34028.

\bibitem{falk1998} M.L. Falk, J.S. Langer, Phys. Rev. E {\bf 57}, 7192 (1998).

\bibitem{Chikkadi2012}V. Chikkadi, S. Mandal, B. Nienhuis, D. Raabe, F. Varnik, P. Schall, Europhys. Lett. {\bf 100}, 56001 (2012). 

\bibitem{lacks2001} D.J. Lacks, Phys. Rev. Lett. {\bf 87}, 225502 (2001).

\bibitem{Regev2015}  I. Regev,	J. Weber,	C. Reichhardt,	KA. Dahmen, T. Lookman, Nat. Commun. {\bf 6}, 9805 (2015).

\bibitem{ediger2012} M.D. Ediger, P. Harrowell, J. Chem. Phys. {\bf 137}, 080901 (2012).

\bibitem{Anderson2002} V.J. Anderson, HNW. Lekkerkerker, Nature {\bf 416}, 811 (2002).

\bibitem{Heussinger2014} R. M\"obius, C. Heussinger, Soft Matter {\bf 10}, 4806 (2014).

\bibitem{Reynaert2007} S. Reynaert, P. Moldenaers, J. Vermant, Phys. Chem. Chem. Phys. {\bf 9}, 6313 (2007).

\bibitem{brooks1999} C.F. Brooks, G.G. Fuller, C.W. Frank, C.R. Robertson, Langmuir {\bf 15}, 2450 (1999).

\bibitem{liu2014} S.S. Schoenholz, A.J. Liu, R.A. Riggleman, J. Rottler, Langmuir {\bf 4}, 031014 (2014).

\bibitem{cande2010} R. Candelier, A. Widmer-Cooper, J.K. Kummerfeld, O. Dauchot, G. Biroli, P. Harrowell, D.R. Reichman, Phys. Rev. Lett. {\bf 105}, 135702 (2010).

\bibitem{khair2016} A.S. Khair, J.Fluid. Mech. {\bf 791}, R5 (2016).


\end{thebibliography}
\end{document}